\documentclass[a4paper]{article}
\usepackage{Odyssey2024}
\usepackage{epsfig,amssymb,amsmath}
\usepackage{multirow}
\usepackage{multicol}
\usepackage{booktabs}
\ninept

\title{1st Place Solution to Odyssey Emotion Recognition Challenge Task1: \\ Tackling Class Imbalance Problem}

\makeatletter
\def\name#1{\gdef\@name{#1\\}}
\makeatother
\name{{\em Mingjie Chen$^1$, Hezhao Zhang$^1$, Yuanchao Li$^2$, Jiachen Luo$^3$, Wen Wu$^4$, Ziyang Ma$^5$,} \\
{\em Peter Bell$^2$, Catherine Lai$^2$, Joshua Reiss$^3$, Lin Wang$^3$, Philip C. Woodland$^4$, Xie Chen$^5$,}\\
{\em  Huy Phan$^{6\star}$, Thomas Hain$^1$}
\thanks{$^\star$The work does not relate to Huy Phan’s position at Amazon.}
}
     
\address{ {\em $^1$ Department of Computer Science, University of Sheffield, United Kingdom } \\
 {\em  $^2$ Centre for Speech Technology Research, University of Edinburgh, United Kingdom} \\
 {\em  $^3$ Centre for Digital Music, Queen Mary University of London, United Kingdom} \\
 {\em  $^4$ Department of Engineering, University of Cambridge, United Kingdom} \\
 {\em  $^5$ Department of Computer Science and Engineering, Shanghai Jiao Tong University, China} \\
{\em  $^6$ Amazon AGI, Cambridge, United Kingdom}\\ 
{\small \tt mingjie.chen@sheffield.ac.uk, t.hain@sheffield.ac.uk}\\
}

\begin{document}
\maketitle

\begin{abstract}
Speech emotion recognition is a challenging classification task with natural emotional speech, especially when the distribution of emotion types is imbalanced in the training and test data. In this case, it is more difficult for a model to learn to separate minority classes, resulting in those sometimes being ignored or frequently misclassified. Previous work has utilised class weighted loss for training, but problems remain as  it sometimes causes over-fitting for minor classes or under-fitting for major classes.
This paper presents the system developed by a multi-site team for the participation in the Odyssey 2024 Emotion Recognition Challenge Track-1. The challenge data has the aforementioned properties and therefore the presented systems aimed to tackle these issues, by introducing focal loss in optimisation when applying class weighted loss. Specifically, the focal loss is further weighted by prior-based class weights. 
Experimental results show that combining these two approaches brings better overall performance, by sacrificing performance on major classes.  The system further employs a majority voting strategy to combine the outputs of an ensemble of 7 models. The models are trained independently, using different acoustic features and loss functions - with the aim to have different properties for different data. Hence these models show different performance preferences on major classes and minor classes. The ensemble system output obtained the best performance in the challenge, ranking top-1 among 68 submissions. It also outperformed all single models in our set. On the Odyssey 2024 Emotion Recognition Challenge Task-1 data the system obtained a Macro-F1 score of 35.69$\%$ and an accuracy of 37.32$\%$.
\end{abstract}

\section{Introduction}
It is commonly assumed that for progress in  conversational AI systems it is essential to enable computers to understand emotions from human speech signals. Speech emotion recognition (SER) is gaining increasing attention due to its wide range of potential application, especially in the context of the recent advancement of large language models \cite{chatgpt_bibtex}. 
SER has been a research focus for a long time, however it is still a complex task because of the multitude of factors that affect the task, including context information, speaking environments, the personality and the speaking style of speakers, language, cultural aspect, commonsense knowledge etc. \cite{el2011survey,madanian2023speech}. 

\begin{figure}[!t]
    \centering
    \includegraphics[height=3in]{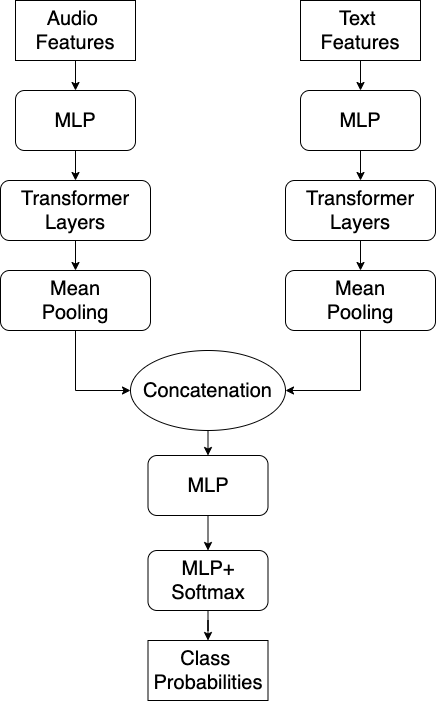}
    \caption{The architecture of each model in the ensemble system. `MLP' denotes multi-layer perceptron.}
    \label{fig:model-arc}
\end{figure}
Typically there are  two  types of SER task due to the annotation style used in emotion labelled datasets, namely classification and regression \cite{madanian2023speech}. In SER classification tasks, speech segments are typically annotated with labels from a small set (4--8) of emotion classes. The task is to predict the correct (single) emotion class representing the complete speech segment.

Many datasets \cite{busso2008iemocap,lotfian2017building,haq2011multimodal} have been created for SER. Most of them \cite{haq2011multimodal} are created by recording actors portraying the required emotion in their speech. Other type of dataset \cite{busso2008iemocap} have been created by prompting speakers to express specific emotions. There are a few datasets \cite{lotfian2017building}, usually referred to as natural datasets,  that are directly collected from sources containing spontaneous speech with natural emotional expressions.
Previous work \cite{milner2019cross,ma2023emotion2vec} has shown that the performance of SER models on these three types of datasets differs significantly. SER classification tasks on natural datasets are still challenging, for  many reasons \cite{gomez2024speech}.

One of the many challenges of SER classification on natural datasets is the imbalanced class distributions. Discriminative machine learning methods typically also choose decision boundaries on the basis of the prevalence of a class. Classes with low occurrence  will not only get a poor representation, but also end up being considered of less relevance. Thus the training of machine learning models is difficult for minor classes. Models can be easily trained on major classes but can tend to ignore minor classes. 
One simple solution is to re-balance (re-weight) the loss of a class by class frequencies.
Many other solutions have been proposed to solve this problem, such as data augmentation \cite{meng2023deep,ma2023leveraging}, new sampling strategies \cite{ibrahim2022bidirectional}, and the use of a modified loss function \cite{yun2009speech}. However, all of the above methods are likely to cause over-fitting problems on minor classes, thus sacrificing performance on major classes. This issue is particularly pertinent here, as the SER system is designed for participation in the Odyssey 2024 Emotion Recognition Challenge Track-1, which has class imbalanced training data, but a class balanced test set. 

We have therefore developed a system that makes use of an ensemble system of 7 models. Each model takes in multi-modal features, from both audio and text. In order to obtain text representations an automatic speech recognition (ASR) model \cite{radford2023robust} is used to generate transcriptions from speech segments. To enhance the quality of the ASR  transcriptions, an error correction model is used in post-processing

In the ensemble system all models share the same architecture. They are trained independently with different audio features or different optimisation configurations. The loss function chosen is  either the focal loss \cite{lin2017focal} or the cross entropy loss, weighted by prior-based class weights or uniform class weights.  

The prior-based class weights are used to give more preference to minor classes than major classes during training.  However, this strategy was found to cause over-fitting for minor classes, and thus a reduction in overall performance.  In order to alleviate this issue, the use of focal loss \cite{lin2017focal} was included, which aims to give higher weights to more difficult samples and lower weights to easier samples.

The model trained with the focal loss together with the prior-based class weights obtained the best overall performance. However, this model performs more poor on major classes than the models trained with uniform class weights. To reach better overall performance the ensemble system is designed to comprise of models with different preferences on major classes and minor classes. Experimental results show that the ensemble system reaches the state-of-the-art performance in the Odyssey 2024 Emotion Recognition Challenge track-1 \cite{Goncalves_2024}. The system obtained a Macro-F1 of 35.69$\%$, and has ranked the first among 68 submissions. 

\section{Related Works}



\subsection{Fusion Techniques for SER}
Over the past decade, research on fusion techniques for SER has made significant progress. Alongside traditional feature-level fusion (i.e., early fusion) and decision-level fusion (i.e., late fusion), there has been widespread exploration of sophisticated tensor-level fusion methods. For instance, \cite{hazarika2020misa} combined both modality-invariant and modality-specific features and applied various regularisation functions to reduce the distance between the modalities. \cite{luo2023mutual} proposed a weighted fusion method based on a cross-attention module for encoding inter-modality relations and selectively capturing effective information. \cite{wu2021emotion} developed a dual-branch model, with one time-synchronous branch that combined speech and text modalities, and a time-asynchronous branch integrating sentence text embeddings from context utterances. \cite{li2022fusing} fused ASR hidden states and ASR transcriptions with audio features in a hierarchical manner.

\subsection{ASR Error Correction}
As outlined above, the system makes use of ASR for transcript generation, using an off-the-shelf system. To enhance ASR performance ASR Error Correction (AEC) methods can be helpful, by post-processing using some knowledge about the task or target domain. The standard method for addressing language domain mismatch is to train an in-domain language model for direct integration with ASR systems. \cite{tanaka2018neural,inaguma2018improving}. However, an alternative is to use AEC sequence to sequence models that correct the output. This is particularly useful in scenarios where the ASR is a black box \cite{liao2023improving}. More recently, there has been interest in employing generative error correction using large language models \cite{radhakrishnan2023whispering}. Furthermore, some studies have explored using both speech and ASR hypotheses as input, instead of relying solely on text data, leading to the development of cross-modal AEC methods \cite{radhakrishnan2023whispering}.

\section{System Description}

As mentioned above, the system yielding the best performance makes use of an ensemble of models. Each model 
takes as input in frame-level audio features as well as token-level text features. The output of each model are the  probabilities for each emotion class. In the majority voting, the prediction of each model is equally used for voting to emotion classes. The most voted emotion class is going to be the final prediction of the ensemble system. 

To extract text features, this work used transcriptions generated by the Whisper-large-v2 model \cite{radford2023robust}. As there are no transcriptions available for the test set, an ASR system is needed for transcriptions, but it will inevitably produce erroneous transcripts. Training an emotion classification model on ground truth transcripts would result in a mismatch between training and test conditions, thus all models make use of ASR output. To enhance the quality of the ASR transcriptions, a sequence-to-sequence ASR error correction model\footnote{https://huggingface.co/YC-Li/Sequence-to-Sequence-ASR-Error-Correction} \cite{li2024crossmodal} is trained and then used to correct errors in the transcriptions in the test set.

\subsection{Model Architecture}

Figure \ref{fig:model-arc} illustrates the principal architecture of models. Frame-level audio features and token-level text features are encoded by two Multi-Layer Perceptron (MLP) modules. Then transformer layers \cite{vaswani2017attention} are used to process audio and text features to encode dynamic information in features. In order to avoid over-fitting, the number of heads in the transformer layers is set to 1. The transformer layers are followed by a mean pooling layer, then the utterance-level audio features and text features are concatenated. The concatenated features are processed by the two MLP modules. The softmax output of the  final MLP produces class probabilities.

\subsection{Audio Features and Text Features}
\begin{table}[!t]
    \centering
    \begin{tabular}{c|c|c|c|c}
    \toprule
     Features    & Modality &  Dim & \#Params & Hours \\
     \midrule
     WavLM-large & Audio & 1024 & 300M & 94K\\
     Hubert-extra-large & Audio & 1280 & 1B &60K\\
     Whisper-large-v3 & Audio & 1280 & 1.5B & 680K\\
       Roberta-large & text & 1024 & 355M & -\\
     \bottomrule
     
    \end{tabular}
    \caption{Audio and text features, where `Dim' denotes the dimensionality of frame-level features, `\#Params' denotes the number of parameters, `Hours' denotes the amount of speech data used for pretraining.}
    \label{tab:features}
\end{table}

The audio and text features used in this work are presented in Table~\ref{tab:features}. Three types of audio features are used: the final layer representations of WavLM-large \cite{chen2022wavlm}; the final layer representations of Hubert-extra-large \cite{hsu2021hubert};  and the final layer output of the encoder of Whisper-large-v3 \cite{radford2023robust}. In terms of text features, this work utilises Roberta-large \cite{liu2019roberta}. In order to enhance the representative capability of the text features, this work uses the average representations of the last 4 layers'  from Roberta-large.
\begin{table}[!t]
    \centering
    \begin{tabular}{c|c|c|c}
    \toprule
       Index & Loss Function & Class Weights & Audio Features  \\
    \midrule
    1& Focal ($\gamma=2$ ) & Prior-based & Whisper \\
   
    2 & Focal ($\gamma= 2.5$ ) & Prior-based & Whisper \\
    3 & CE & Prior-based & Whisper \\
    4 & Focal ($\gamma=2$) & Uniform & Whisper\\
    5 & CE & Uniform & Whisper \\
    6 & Focal ($\gamma=2$) & Prior-based & WavLM \\
    7 & Focal ($\gamma=3$ ) & Prior-based & Hubert \\
    \bottomrule
    
    \end{tabular}
    \caption{Configurations of models in the ensemble systems, where `CE' denotes cross-entropy. }
    \label{tab:ensemble_models}
\end{table}

\begin{table*}[!t]
    \centering
    \begin{tabular}{c|c|c|c||c|c|c}
    \toprule
     Subset  & \# Samples All & \# Speakers All & Duration All & \# Samples Used & \# Speakers Used & Duration Used  \\
     \midrule
      Training   & 68119 & 1405 &110.2h  &53386 &1391 &86.3h  \\
      Dev & 19815 &454 &31.7h &15341 & 446 & 24.4h  \\
      Test-3 & 2437 &187  &3.9h & - & - & - \\
     \bottomrule
    \end{tabular}
    \caption{The detailed information of subsets of the MSP-Podcast v1.11, where `All' denotes subset statistics before the filtering and `Used' denotes subset statistics after the filtering.}
    \label{tab:subsets}
\end{table*}

\begin{table}[!t]
    \centering
    \begin{tabular}{c|c|c}
    \toprule
    Class  & \# Training Samples & \# Dev Samples   \\
    \midrule
    Neutral & 25016 &5667 \\
    Happy & 13440 & 3340\\
    Angry & 3053 & 2413\\
    Sad & 3882 & 1101\\
    Disgust & 1426 & 486\\
    Contempt & 2443 & 1323\\
    Surprise & 2897 & 729\\
    Fear & 1139  & 282\\
    \bottomrule
    
    \end{tabular}
    \caption{Class distributions in the training set and the development set after the filtering.}
    \label{tab:class_distribution}
\end{table}

\subsection{Loss Functions and Class Weights}
This work considers two options for loss functions, the focal loss and the cross entropy loss. The loss function can be weighted by the prior-based class weights or uniform class weights. Combining loss functions with class weights, four types of optimisation configurations can be used.

\subsubsection{Uniform Class Weights and Prior-based Class weights}
Consider a classification task with $C$ classes, a training dataset $\{(\mathbf{x}_i, y_i) | i = 1,...,N\}$, $y_i \in \{1,..., C\}$, $\mathbf{x}_i \in \mathrm{R}^{d}$. The uniform class weights can be written as follows,
\begin{equation*}
    \mathbf{w}_\text{uni} = \{w_j = 1| j = 1,...,C\}.
\end{equation*}
Prior-based class weights can be written as:
\begin{equation*}
    \mathbf{w}_\text{prior} = \{w_j = \frac{N}{N_j} | j = 1,...,C\},
\end{equation*}
where $N_j$ is the number of samples in class $j$ in a training or development set.

\subsubsection{Cross-Entropy Loss and Focal Loss}

Following the above notations , the cross-entropy loss can be defined as:
\begin{equation*}
    \mathcal{L}_\text{ce} = 1/N \sum_{i=1}^{N} - \log p_i,
\end{equation*}
where $p_i = P(y_i | \mathbf{x}_i)$ is the output probability of the corrected class.
The focal loss can be written as follows,
\begin{equation*}
    \mathcal{L}_\text{focal} = 1/N \sum_{i=1}^{N} - (1-p_i)^{\gamma} \log p_i,
\end{equation*}
where $\gamma$ is a hyper-parameter. 

By combining with the class weights, the class-weighted focal loss can be written as:
\begin{equation*}
    \mathcal{L}_\text{focal}' = 1/N \sum_{i=1}^{N} - w_j (1-p_i)^{\gamma} \log p_i,
\end{equation*}
where $w_j$ is the class weight for class $y_i$.

\subsection{Ensemble Strategy}

Table \ref{tab:ensemble_models} lists the configurations of the 7 models in the ensemble system. All models use the averaged representations of the last 4 layers of Roberta-large (refer to Table \ref{tab:features}) as text features. Apart from the optimisation configurations, the values of the focal loss hyper-parameter was explored with $\gamma = 2$ and $\gamma = 2.5$, as well as three types of audio features.

\section{Experimental Setup}

\subsection{Dataset}
The Odyssey 2024 Emotion Recognition Challenge \cite{Goncalves_2024} used the MSP-Podcast dataset v1.11 \cite{lotfian2017building}. The dataset is derived from podcasts, and annotated through crowd-sourcing. Different from most datasets containing acted speech \cite{haq2011multimodal} or elicited speech \cite{busso2008iemocap}, this dataset contains spontaneous speech with natural human emotions. The dataset is composed of five subsets: the training set, the development set, the test-1 set, the test-2 set, and the test-3 set. In this challenge, the test-3 set is used to measure the outcome, and the reference labels have not been made public.
The speech segments have been annotated with 10 classes, in which 8 classes are used for this challenge. These are Neutral, Happy, Angry, Disgust, Sad, Surprise, Contempt and Fear. There are two remaining labelled classes, Other and No Agreement, are not used.  Since the Other and No Agreement classes are not used in the challenge, this work removed associated samples, retaining only samples with the challenge classes.
Detailed information of the training set, development set and test set are given in Table \ref{tab:subsets}, including the subset statistics, before and after the sample removal.

\subsection{Implementations}
\subsubsection{Multi-Modal Classifier Model}
The Hubert-extra-large model, the WavLM-large model, the Whisper-large-v3 model and the Roberta-large model are imported from the transformers toolkit \cite{transformers}. 

In terms of the model architecture, the hidden size of the transformer layers is 512 and the number of the transformer layers is set to 2. 
The MLP module before the transformer layers has a hidden size of 512. The MLP modules after the concatenation layer has a hidden size of 512 and the output size of the final MLP module is 8.

The models are trained with a batch size of 128, an initial learning rate of 1e-4, with a learning scheduler \cite{su2015experiments}. 
The model checkpoint of the epoch with the best Macro-F1 on the development set is chosen for evaluation.
The implementations are based on the recipe of the speechbrain toolkit \cite{ravanelli2021speechbrain}. All models are implemented and trained with the PyTorch toolkit \cite{paszke2019pytorch}. Training takes about 2 hours and uses a maximum of 20 epochs.

\subsubsection{Feature Fusion}
A number of methods for feature fusion were considered. Due to limited time, however, only the following ones were implemented and compared. \\
\textbullet\ \textit{Early fusion}: text and audio features are concatenated at the embedding level. \\
\textbullet\ \textit{Late fusion}: text and audio features are learned independently and the final decision is determined based on respective outputs. \\
\textbullet\ \textit{Early fusion} + \textit{late fusion}. \\
\textbullet\ \textit{Tensor fusion}: unimodal information and bimodal interactions are learned explicitly and aggregated \cite{zadeh2017tensor}. \\
\textbullet\ \textit{Low-rank tensor fusion}: multimodal fusion with modality-specific low-rank factors, which
scale linearly in the number of modalities \cite{liu2018efficient}.

While sophisticated fusion approaches have often outperformed early fusion in various scenarios, this phenomenon was not observed in the experiments. Therefore,  early fusion was used in the model, as it was found to outperform all other methods tested.
 
\subsubsection{ASR Error Correction}
A pretrained AEC model was used in this work, which has been trained on the English version of Common Voice 13.0 \cite{ardila2019common} and TED Talk corpus \cite{cettolo2012wit3} using a publicly available Sequence-to-Sequence (S2S) encoder-decoder architecture \cite{chen2023exploring}. This model was trained to convert ASR transcriptions to human-transcribed transcriptions (i.e., ground-truth text). Considering the disparity between the MSP-Podcast dataset and the two AEC pretraining corpora (e.g., out-of-domain words), this model was fine-tuned to enhance its performance. Specifically, the model was trained on the training set of the provided MSP-Podcast corpus for 10 epochs and then validated on the development set. The best checkpoint was saved to correct errors on the test set. The correction quality was evaluated using WER, BLEU, and GLEU scores for a comprehensive assessment. The results of the best checkpoint on the development set are presented in Table \ref{tab:AEC}, demonstrating the effectiveness of the AEC model. As there is no ground-truth text for the test set, there is no way to further evaluate its effectiveness. Given the same domain of the development and test sets, it is expected an improvement in accuracy on the test set by approximately 1\%, as suggested by previous research on the impact of WER on SER performance \cite{li2023asr}.

\begin{table}[ht!]
\centering
\begin{tabular}{c|c|c|c}
\toprule
Transcription & WER $\downarrow$ & BLEU $\uparrow$ & GLEU  $\uparrow$ \\
\midrule
Original & 17.65 & 81.32 & 78.02 \\ 
Corrected & 14.51 & 83.48 & 81.19 \\
\bottomrule
\end{tabular}
\caption{Comparison of the quality between the original and corrected transcriptions on the development set. All values are presented in percentage scale (\%). $\downarrow$: the lower the better. $\uparrow$: the higher the better.}
\label{tab:AEC}
\end{table}

\subsection{Evaluation Metrics}
According to the challenge evaluation setup, Macro-F1 is used as the primary metric. Macro-F1 is the unweighted average of the F1 score of each class. Apart from Macro-F1, the weighted accuracy (WA) and unweighted accuracy (UA) are used. \footnote{The accuracy\_score function and the balanced\_accuracy\_score function from the scikit-learn toolkit are used for implementing unweighted accuracy and weighted accuracy, respectively}

\section{Results}

\begin{table*}[!t]
    \centering
    \begin{tabular}{c|c|c|c|c|c|c|c}
    \toprule
     Index & Audio & Text & Loss & Class Weights & Macro-F1 ($\%$) & WA ($\%$) & UA ($\%$) \\
     \midrule
     1 & Whisper & Roberta & Focal ($\gamma=2$) & Prior-based & \textbf{34.2} & 35.6 & 45.7  \\
     2 & WavLM & Roberta & Focal ($\gamma=2$) & Prior-based & 32.4 & 35.3 & 43.6\\
     3 & Hubert & Roberta & Focal ($\gamma=2$) & Prior-based & 32.7 & 34.3 & 45.5\\
     4 & Whisper & Roberta & Focal ($\gamma=2.5$) & Prior-based & 33.8 & 35.8 & 45.1   \\
     5 & Whisper & Roberta & CE & Prior-based &  33.8 & \textbf{36.0} & 44.0  \\
     6 & Whisper & Roberta & Focal ($\gamma=2$) & Uniform & 33.3 & 32.9 & \textbf{51.9}    \\
     7 & Whisper & Roberta & CE & Uniform & 32.8 &32.6 & 51.1   \\
     \midrule
     Ensemble & - & -& -& -& \textbf{35.6} & \textbf{36.6} & 49.3 \\
     \bottomrule
    \end{tabular}
    \caption{Development set results of the 7 models, they are trained independently with different audio features, text features, loss and class weights.}
    \label{tab:results}
\end{table*}

Results for the 7 models in the ensemble system are presented in Table \ref{tab:results}. Generally speaking, different configurations of audio features, loss functions and class weights yield differences in WA and UA results. 
From analysing the performance difference in WA and UA results, it would be easy to understand how the models perform on the major classes and minor classes.  
For convenience, in the following discussion, the index of the models in Table \ref{tab:results} are used to denote the models (e.g. model-1). 

Among the 7 models, the best Macro-F1 was obtained by the model-1, applying the focal loss ($\gamma=2$) and prior-based class weights. Comparing model-1 to model-4, increasing $\gamma=2$ causes an improvement in WA but a drop in UA, causing a reduction in Macro-F1.  

Among the 3 types of audio features for model-1, model-2 and model-3, it is clear that the Whisper features yield better results than both the WavLM features and the Hubert features. One possible reason is that the Whisper model were trained supervisedly with text transcriptions, while the other two models were trained without supervision. These are also difference due to the model sizes and the amount of training data used bay these pretrained models.  

The results for model-1, model-5, model-6 and model-7 can help to understand the effect of the two loss functions and the two class weights. Generally speaking, different loss functions and class weights yields models that have different preferences for major and minor classes.
Specifically, when training with the uniform class weights, model-6 and the model-7 show good performance on UA but poor performance on WA. Comparing model-7 and model-5, the prior-based class weights give more attention to minor classes, causing a significant improvement in WA but a large drop in UA. This means that the prior-based class weights improve the performance on the minor classes, but sacrificing the performance on the major classes. The performance drop may be due to model-5 over-fitting on major classes. The effect of focal loss can be found through comparing model-1 and model-5, which shows that focal loss helps model-1 reach a better balance between the major classes and the minor classes. Hence, the model-1 reaches the highest overall macro-F1 performance. 

Based on the diverse performance of the models, an obvious strategy to build an ensemble is to combine the outputs of these models through a majority voting process. The results show that the ensemble system outperforms all of the 7 models on Macro-F1 and WA, reaching the state-of-the-art performance.

\section{Conclusions}
This paper introduces an ensemble system that includes 7 multi-modal models, constructed for participation in the Odyssey 2024 Emotion Recognition Challenge. The system showed the best performance among a total of 68 submissions to the challenge, in all metrics under consideration.  
The 7 models were trained independently with different loss functions and class weights. Specifically, the cross entropy loss and the focal loss were used. Uniform class weights and prior-based class weights are studied. The experiment results show that the combinations of loss functions and class weights lead to different preference on the major classes or the minor classes

\section{Acknowledgements}
This work was primarily conducted at the Voicebase/Liveperson Centre of Speech and Language Technology at the University of Sheffield which is supported by Liveperson, Inc..

\vfill\pagebreak

\bibliographystyle{IEEEbib}
\bibliography{Odyssey2024_BibEntries}

%

\end{document}